\input epsf
\documentstyle[twocolumn,aps,epsf]{revtex}
\begin{document} 
\twocolumn[\hsize\textwidth\columnwidth\hsize\csname 
@twocolumnfalse\endcsname
%\draft{} 
%\preprint{}
%\receipt{} 
\title{Size, Shape and Low Energy Electronic Structure of Carbon Nanotubes}
\author{C.L. Kane and E.J. Mele}
\address{ Department of Physics \\
  Laboratory for Research on
the Structure of Matter\\ University of Pennsylvania \\ Philadelphia, 
Pennsylvania 19104}
\maketitle
\begin{abstract}
 A theory of the long wavelength low energy electronic structure of graphite-derived
 nanotubules is presented. The propagating $\pi$ electrons are described by wrapping a massless
 two dimensional Dirac Hamiltonian onto a curved surface. The effects of the tubule size, shape and
 symmetry are included through an effective vector potential which we derive for this model. The rich gap structure for all
 straight single wall cylindrical tubes is obtained  analytically in this theory, and the
 effects of inhomogeneous   shape deformations  on  nominally metallic armchair  tubes are analyzed. 
\end{abstract}  
]
 Since the discovery of a new  family of carbon based structures formed by folding
 graphite sheets into compact tube-shaped objects, there has been
 interest in the electronic properties which can be realized with these
 structures \cite{review}.  It is now understood that these tubes exhibit insulating,
 semimetallic or metallic behavior depending on  the  helicity of the mapping of the graphite sheet onto the surface of the
 tube \cite{Saito,Ham,Mint1}. Discrete microscopic defects, in the form of disclination pairs, 
 provide an interface between neighboring straight tubule segments of
 \it different \rm helicities with different electronic gaps, providing  a novel
 class of elemental heterojunctions \cite{Chico}. 

 In this Letter we investigate the effects of shape fluctuations on the electronic 
 properties of the carbon nanotubes. We present a new formulation
 of this problem which allows us to study the effects of geometry on the quantum dynamics for
 a $\pi$ electron propagating 
 within the surface of the wrapped graphite sheet. We show that the very rich
 phenomenology already well established for straight single wall cylindrical tubules can be derived directly 
 from this geometrical theory \cite{Saito,Ham,Mint1}. We then extend the model to consider the effects of \it inhomogeneous
  \rm deformations in the form of local twists and bends of the tubule on the low energy
 electronic structure. These are important low energy structural degrees of freedom of 
the tubules, and indeed  one finds that these deformations are easily
 quenched into any three dimensional network composed of tubules. 
  We show that these shape fluctuations also have a very strong effect on the low energy electronic and transport properties.

 An isolated two dimensional sheet of graphite is a semimetal, with the Fermi energy
 residing at a critical point in the two dimensional $\pi$ electron spectrum. 
 The Fermi surface is collapsed to a point for this system; there are  two distinct
 Fermi points at $K$ ($K'$) points of the zone $ (\pm 4 \pi/3 a,0)$, where $a$
is the length of the primitive translation vector
($a = \sqrt{3} d$ where d is the nearest neighbor bond length of the graphite lattice). 
Expanding the $\pi$ electron Hamiltonian around either of these points,  and linearizing in spatial
 derivatives, one finds that the low energy electronic states are described by a massless two dimensional Dirac Hamiltonian,  
   $ H_{\rm eff} = v \vec \sigma  \cdot \vec p $, 
 where $p$ denotes a two dimensional momentum in the graphite plane, and the $\sigma$'s are the
 2 $\times$ 2 Pauli matrices \cite{dave}. Here the two spin polarizations of the particle refer to the two independent basis  states (labeling the a and b sublattices) in the graphite primitive cell. Thus, in 
addition to its physical spin and momentum, the  $\pi$ electron carries an internal pseudospin
 index, labeling the sublattice state, and an isospin index, labeling the two independent
 Dirac spectra  derived from the K and K' points of the zone. The Fermi energy for this system is at E=0. 

 To study the electronic behavior of the tubule one maps $H_{\rm eff}$  
 onto a curved surface.  The mapping of the graphite sheet onto the
 cylindrical surface can be specified by a single superlattice translation vector $T_h$ which
 defines an elementary orbit around the waist of the cylinder. In the absence of disclinations
 the superlattice vector $T_h$ is an element of the original graphite triangular Bravais lattice.  This wrapping is conventionally indexed by two integers
 [$m,n$] such that $T_h$ = $m T_1 + n T_2 $ with $T_1 = a(1,0)$ and  $T_2 = 
 a(1/2,\sqrt{3}/2)$ \cite {Saito,Ham,Mint1}.   In figure (1)  we show the structures of three cylindrical single
 wall nanotubes for selected values of  $m$ and $n$. The 
 $\pi$ electron eigenstates $\Psi(r)$ satisfy periodic boundary conditions on the
 cylinder, i.e. $\Psi(r+T_h) = \Psi(r)$.  Interestingly this does \it not \rm imply 
 periodic boundary conditions for the eigenfunctions of $H_{\rm eff}$.
 Instead,  $H_{\rm eff}$ is obtained from a factorization of the single particle
 state $\Psi = \psi(r) \cdot (U_a(r) , U_b(r)) $
 where  $(U_a , U_b)$  are the two sublattice components of the eigenstates of the $\pi$ electron Hamiltonian at the
 critical K point of the zone, and $\psi (r)$ is  a (slowly varying) eigenstate of  $H_{\rm eff}$
 . In particular, the functions $U_a$
 and $U_b$ are Bloch functions with crystal momentum $K = (4 \pi/3a,0)$ which do not
 retain the periodicity of the original Bravais lattice, but are invariant only under
 the translations of a $\sqrt{3} \times \sqrt{3}$  superlattice.  Since $\Psi(r)$ is
 invariant under Bravais lattice vector $T_h$, one finds that the function $\psi$ is 
 conjugate to $U$, and accumulates a phase $e^{-i K \cdot T_h}$ under translations $T_h$.
 
 This phase shifted boundary condition is an awkward computational as well as conceptual
 constraint on the low energy theory. However, we observe that this constraint can
 always be enforced by imposing strictly periodic boundary conditions on the wavefunction
 $\psi$ in the presence of an effective vector potential $a_w$ which satisfies
  $2 \pi \int  a_w\cdot dl = - K \cdot T_h$
 where the line integral is taken on a closed orbit around the waist of the cylinder. 
 This vector potential  can be  associated with an elementary flux of strength
 $\Phi = - K\cdot~T_h / 2 \pi$ which links  the cylinder. 

 The azimuthal quantum states which satisfy periodic boundary conditions on the surface 
 of the cylinder are the cylindrical harmonics $e^{i m \phi}$ with integer $m$.  We adopt a 
 coordinate system in which the $\xi$-direction denotes the (counterclockwise) tangential direction on the tube's
 surface, and the $\zeta$ direction is aligned along the tube.  Then for the
 $m$-th channel propagating along the tubule
 \begin{equation}
 H_{\rm eff} = \frac{2 \pi v}{T_h} \sigma_1 (m + \Phi)  - i  v \sigma_2 \partial_\zeta
 \end{equation} 
 with spectrum $E = \pm v \sqrt{ q_\zeta^2  + (2 \pi/T_h)^2(m+ \Phi)^2}$.
 Thus, for any wrapping where $\Phi$ is an integer, the accumulated phase due to the vector potential
 can be absorbed into  the definition of the  azimuthal quantum number $m$. In particular, there exists an azimuthal
 state $m = -\Phi$ for which the ``mass" term vanishes, and the electronic spectrum is 
 gapless.  This occurs for the wrapped tubules in  which $T_h$ is an element
 of a regular $\sqrt{3} \times \sqrt{3}$ superlattice of the original graphite Bravais
 lattice. For the remaining two-thirds
 of the wrapped structures, the minimum value of $|m + \Phi| = 1/3$, so that these
 structures retain a nonvanishing gap $\Delta E =   4 \pi v/ 3 T_h$. In a nearest neighbor
 tight binding model, with nearest neighbor hopping 
 amplitude t, one has  $v =  3td /2$ so that these
 primary gaps depend inversely on the tube  radius R, $\Delta E = td/R$
 as has already been deduced from numerical work by several groups. It is 
 noteworthy that the symmetry of the Dirac spectrum requires that for
 the metallic tubules for each isospin there is  only a single azimuthal branch  of the electronic spectrum which
 crosses the Fermi energy,
 \it independent  \rm of the  radius of the tubule, and that transport in a single
 tube is therefore always governed by a single transverse channel. 

In addition to the wrapping constraint described above, the local {\it shape} of the
tubule plays an essential role in determining the low energy electronic properties.
Specifically, curvature and shear in a graphite sheet introduce variations in the
local electronic hopping amplitudes.  Consider
 fluctuations in the hopping amplitudes $\delta t_a$ along three nearest neighbor bonds $\vec\tau_a$ at a
 given site on sublattice a.  The average value of $\delta t_a$ simply renormalizes the velocity of
the Dirac particle.  However, the variation from bond to bond introduces a new symmetry
breaking term into $H_{\rm eff}$.  Inserting $\delta t_a$ into
the lattice Hamiltonian and expanding about $K$,  this term has the
form $\delta H_{\rm eff} = (v/2)(a_c^+ \sigma^- + a_c^- \sigma^+)$, where
\begin{equation}
a_c^\pm = {1\over v} \sum_{a=1}^3 \delta t_a e^{\pm i \vec K \cdot \vec\tau_a}
\end{equation}
and $\sigma^\pm = \sigma_1 \pm i \sigma_2$.  Defining the vector
\begin{equation}
\vec h =  \sum_{a=1}^3 \hat \tau_a \delta t_a/t .
\end{equation}
it is straightforward to show $a_c^\pm = i (2/3d)(h_x \pm i h_y)$.
This term may be written as an effective
curvature derived vector potential \cite{connection}$\vec a_c  = \hat z \times\vec h$,
where $\hat z$ is the  local unit normal vector.
We then have,
 \begin{equation}
 H_{\rm eff} = v \vec\sigma \cdot (\vec p + \vec a_w + \vec a_c)
 \end{equation}
A similar expression may be derived for the $K'$ point.
 Equation (4) demonstrates that the effect of fluctuations in the bond hopping amplitudes
 are to displace the singular point of the Dirac operator in k space, but not
 to remove it. As an electron propagates on the surface of a tube it accumulates
 a phase from both the winding condition (through $a_w$) and from the local fluctuations
 of the hopping amplitudes (through $a_c$) which it encounters along its path. 

 The bond hopping fluctuations can now be deduced from the shape of the tubule. Here 
 we consider contributions from bond length variations, and from the misorientation
 of $\pi$ electron orbitals on neighboring sites of the tubule.
  The detailed calculations leading to the results given 
 below are straightforward but lengthy \cite{KM}, and will not be presented here. Instead
 we focus on the key results. One finds
 that the dependence of the hopping amplitudes on bond length for a hop
 along bond $\vec\tau_a$ can be expressed in
 terms of the metric tensor $g_{ij}$ for the curved surface so that
 \begin{equation}
 \delta t_a/t = (\beta/2d^2) \tau_a^i \tau_a^j(g_{ij} - \delta_{ij})
 \end{equation} 
 where 
 $\beta = \partial \ln t / \partial \ln d$ gives the linear dependence of the bond
 hopping operator on bond length. We consider the $\pi$ orbitals to be oriented along
 the local normal of the tubule surface. On a curved surface, the local normals on
 two neighboring sites are no longer perfectly aligned, and this misorientation effect also modulates
 the hopping amplitudes. We find that this effect can be calculated using both the metric tensor
 $g_{ij}$ 
 and the curvature tensor $K_{ij}$ \cite{fdavid}.  The result is
 \begin{equation}
 \delta t_a/t  = - (1/8) \tau_a^i \tau_a^j  K_{jk} K_{il} g^{kl} + ... 
 \end{equation}
 There is an additional contribution which arises  from rehybridization of the $\pi$ electron
 states on the curved manifold, and it can be derived by studying the effects of shape
 fluctuations on the invariant $(\hat{n} \cdot \vec\tau_a) (\hat{n}' \cdot \vec\tau_a)$ \cite{KM}. The essential
 ingredients for the long wavelength physics are already contained in the  former two contributions, 
 and  we now consider them in more detail. 
 
  For a tubule in the form of a right circular cylinder we need to specify the metric tensor, 
 curvature tensor, and the tipping angle $\theta$ which orients a bond of the honeycomb network with
 respect to the ``$\zeta$ axis" along the length of the tube.  For the right circular
 cylinder we have
 $g_{ij} = \delta_{ij}$ and the only nonzero component of the curvature
 tensor is $K_{\xi\xi} = 1/R$.   
 Thus the pure metric contribution to $\vec a_c$ vanishes, and we only  have the orientational contribution.
 The explicit form may be found using (3) and (6) along with the fact that for any three vectors $\vec A$, $\vec B$,
$\vec C$, $\sum_a (\hat\tau_a\cdot\vec A)(\hat\tau_a\cdot\vec B)(\hat\tau_a\cdot\vec C) = (3/4) {\rm Re}
[A^+ B^+ C^+$].  Expressing $\vec a_c$ in the new coordinate system, we find
$a_c^+ = a_{c\xi}+i a_{c\zeta} = - e^{3 i \theta} d /16R^2$.
 Note that this contribution is proportional to the square of the tubule curvature, and
 is unchanged under rotations of the tipping angle by $2 \pi/3$ as one 
 expects from the symmetry of the honeycomb lattice. For the ``zig-zag" tubes  (as shown in 
 figure 1(a))  we have
 a bond exactly aligned with the long axis of the cylinder, so that $\theta = 0$ and $a_c^+$
 is purely real.  This  means that $\vec a_c$ is directed  along the circumferential direction of the tube.
 Here the line integral $\int a_c \cdot dl$ around the waist of the tube is nonvanishing (and
 in general nonintegral), so that the curvature makes a nonzero 
 contribution to the effective mass of the Dirac particle in equation (4).  For the armchair tubes (as shown in figure 1(c))  $\theta = 
 \pi/2$ so that the vector potential is purely imaginary, indicating that it is 
 directed exactly \it along \rm the tube direction. For this geometry its
 only effect to rigidly shift the electronic spectrum along the $q_\zeta$ direction in momentum space. This
 shift has no physical consequence and can be completely eliminated from this geometry by a simple
 gauge transformation. Since $\Phi$ is an integer for all
 armchair tubes, and the curvature corrections can be removed from the Hamiltonian 
by a gauge transformation, 
  the  mass term vanishes and  all the straight armchair tubules remain metallic.

 For a given cylindrical tubule the total band gap is $|2 v (a_w +  {\rm Re}[a_c^+])|$, and 
 in figure 2 we display a plot of the total gaps predicted for all tubules with radii less than
  15 A.  The plot identifies three distinct families of tubules: (a) tubes with primary winding induced gaps
 scaling  as $ 1/R $ with curvature-derived fluctuations scaling as $1/R^2$, 
 (b)  tubes with vanishing primary gaps, and a nonvanishing curvature induced gap (these are
 shown on an expanded scale in the lower panel), and (c) zero gap (armchair tubules) for
 which both the primary gaps and curvature induced gaps vanish by symmetry.  The data
 predicted within this model provide a strikingly complete description of numerical data for these
 gaps
 obtained from a complete tight binding analysis of these tubes employing four basis orbitals
 per carbon site for each of these structures \cite{Mint2}.  Thus the Dirac model on the curved surface
 correctly  represents the low energy electronic physics for this system.  We remark that the scales of these gaps for tubes of radius $\approx$ 10A are by no means negligible and can have an 
 important consequences for the low temperature transport properties \cite{Blase}. 

 This model can now be extended to far more complex structures which contain fluctuations
 in the tubule shape.
 Physically, just as the uniform vector potential describes a homogeneous
 mapping of the graphite plane onto the tubule surface, a perturbation to  the tubule  shape
 produces a perturbation in the  vector potential which then can scatter a quantum particle. Here we 
 will focus only the armchair tubes, since these are the only structures which
 are metallic in the absence of shape fluctuations.  We consider the effects of
 long wavelength twists and bends of the tubule, as shown in figure (3) since these are the low energy
 degrees of freedom for the system.

 We find that even a modest twist can serve as a strong scatterer for a propagating $\pi$ electron.
 The dominant effects are introduced through the metric tensor contributions in 
 equation (5). For a tube subject to a twist 
$\gamma = \partial_\zeta \phi$ we find
$
  a_c^+ =  ie^{3 i \theta} \beta(R/2d)\gamma.
$
 For the armchair tube, $\vec a_c$ is directed along the circumferential direction and
 provides a gap in the Dirac spectrum of $(3\beta \gamma R/2) t$. 
 Physically this effect is due to an asymmetric compression and dilation of the ``axial" bonds on the
 surface of the twisted
 tubule.  
Here we find that
 a  twist which rotates the wrapped graphite structure through an
 angle of $\pi$ over a distance of $ 1 \mu$ introduces a gap of  20 meV  at the Fermi surface. 
The contributions from the curvature induced misorientation of the $\pi$ orbitals
 in equation (10) are considerably smaller by a factor $d^2/12\beta R^2$.

Interestingly, we find that the coupling to bend is much weaker.
A deformation with a constant bend but no twist does not backscatter a particle. 
The underlying reason for this is that an armchair tube with uniform bend has a local mirror plane  
which preserves the symmetry between the ``axial" bonds.  This ensures that the effective vector
potential points along the tube, so it is ineffective for backscattering a propagating particle.
In principle, a propagating $\pi$ electron can scatter from a spatially
 variation of the bend. However, the  first order coupling to a spatial
 derivative of the curvature $\partial_{\zeta} K_{\zeta \zeta}$\cite{curvature} vanishes
 due to the azimuthal symmetry of the electronic states at the Fermi energy.  From
the higher order corrections we estimate a bend induced gap of order
$t R^3 d / \Lambda^4$, where $\Lambda$ is the persistence length of the tube.
This is of order 0.01 meV, which is
negligible for the situation of experimental interest.

 In a real single wall tubule, one expects that the twist can be inhomogeneously
 distributed along the tubule length. (In general, one expects no twist for 
 an isolated  nonchiral tubule, but the torques on a tubule induced by three dimensional packing of
 these structures undoubtedly induced some degree inhomogeneous twist.) For a twisted
 section of tubule connecting two untwisted armchair tubes, one can regard the
 connecting segment as a weak link between conducting segments. At sufficiently low
 temperature, backscattering from these defects can  ultimately lead to localization
 of a quantum particle. However, before this idea can be meaningfully applied to these
 systems, the model will have to be generalized to describe the competing effect of interwall
 quantum coherence  in  three dimensional samples  built out of single wall tubes. Nevertheless
 we remark that recent measurements on ropes composed of armchair tubes, and on mats
 composed of an ensemble of connected ropes, show a resistivity which crosses over
 from a low temperature regime with resistivity decreasing with increasing temperature to
 a high temperature regime in which the resistivity is increasing roughly linearly with
 temperature \cite{jack}.  This crossover occurs in the range 10K - 200K (depending on sample
 quality and morphology) and since these temperatures are in an energy  range which can be
 easily accounted for by the shape fluctuations discussed above it is tempting to associate 
 this crossover with the onset of strong backscattering from quenched disorder in the
 tubule twist. It would be quite interesting to quantify this idea by 
measuring the degree
 of twist which is  actually quenched into three dimensional samples composed of carbon nanotubes.  The
 transport properties of nanotubes subject to a controlled, but variable, torsion
 would similarly provide important information about these effects. Finally, we
 remark that the twist  can be thermally excited and provides an important
 temperature dependent scattering rate for $\pi$ electrons propagating along the tubule. 

 It is a pleasure to thank J. E. Fischer and R. Kamien for helpful discussions.
 This work was supported by DOE under grant DE-FG02-84ER45118, and by
 the NSF under grants DMR 93 13047 and DMR 95 05425.

\begin{figure}
\epsfxsize=3in
\centerline{\epsffile{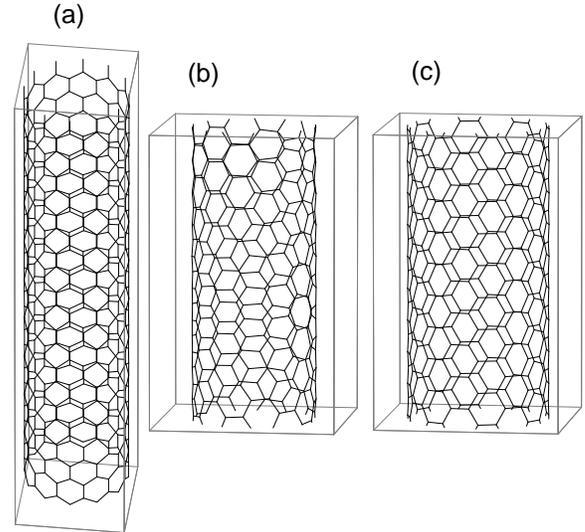}}
\caption{Lattice structures for:  (a) the ideal zigzag [12,0] tube, (b)
 the ideal chiral [8,6] tube, and (c) the ideal armchair [8,8] tube} 
\end{figure}
\begin{figure}
\epsfxsize=3in
\centerline{\epsffile{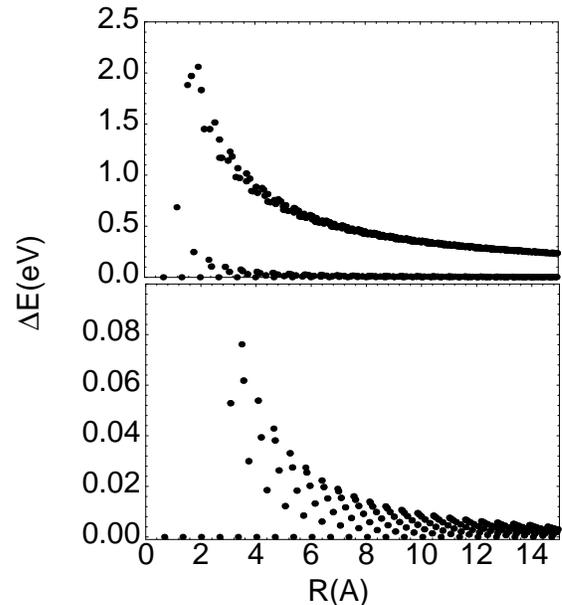}}
\caption{Gaps calculated for all right circular tubes with radii less
 than 15 A. The tubes fall into three families: those with primary gaps
 which scale as 1/R (top panel, top curve), those with zero primary
 gap but nonzero curvature induced gaps which scale as ${\rm 1/R^2}$ (lower
 curve top panel, and shown in the expanded scale in the lower panel)
 and armchair tubes with zero primary gap and zero curvature induced
 gap. } 
\end{figure}

\begin{figure}
\epsfxsize=3in
\centerline{\epsffile{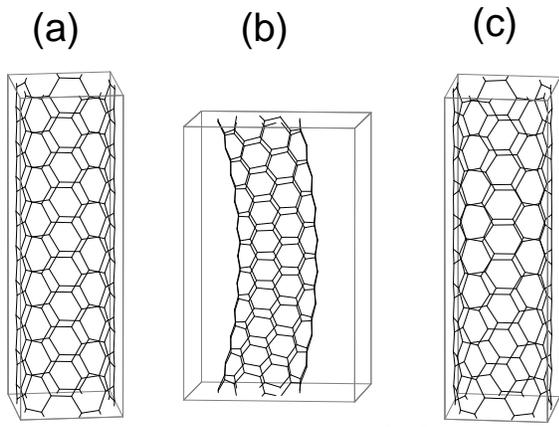}}
\caption{Deformations of a armchair [5,5] tube. The left panel gives
 the ideal tube structure. The middle and right hand panels show the
 effects of uniform bend and twist on the structure.}
\end{figure}


\begin{references}
\bibitem{connection}  $\vec a_c$ is not the related to the spin connection for the curved surface,
whose curl is the Gaussian curvature.
\end{references}
\end{document}